\pdfoutput=1
\RequirePackage{lineno} 
\documentclass[aps,prl,twocolumn,showpacs,superscriptaddress,groupedaddress]{revtex4}  
\usepackage{graphicx}  
\usepackage{dcolumn}   
\usepackage{bm}        
\usepackage{amssymb}   
\usepackage{epstopdf}
\usepackage{epsfig,pslatex,subfigure}

\def \met {$ \hspace{0.02in} / \! \! \! \! \! E_{T} \hspace{0.02in}$}
\def \metmath { \hspace{0.02in} / \! \! \! \!\! E_{T} \hspace{0.005in} }
\def \metb {$ \hspace{0.01in} / \! \! \! \! \! E_{T} \hspace{0.00in}$}

\begin{document}

\hspace{5.2in} \mbox{Fermilab-Pub-10/301-E}

\title{Search for diphoton events with large missing transverse energy in 6.3 fb$^{-1}$ of $\mathbf{p\bar{p}}$ collisions at $\mathbf{\sqrt{s}=1.96}$~TeV}
\affiliation{Universidad de Buenos Aires, Buenos Aires, Argentina}
\affiliation{LAFEX, Centro Brasileiro de Pesquisas F{\'\i}sicas, Rio de Janeiro, Brazil}
\affiliation{Universidade do Estado do Rio de Janeiro, Rio de Janeiro, Brazil}
\affiliation{Universidade Federal do ABC, Santo Andr\'e, Brazil}
\affiliation{Instituto de F\'{\i}sica Te\'orica, Universidade Estadual Paulista, S\~ao Paulo, Brazil}
\affiliation{Simon Fraser University, Vancouver, British Columbia, and York University, Toronto, Ontario, Canada}
\affiliation{University of Science and Technology of China, Hefei, People's Republic of China}
\affiliation{Universidad de los Andes, Bogot\'{a}, Colombia}
\affiliation{Charles University, Faculty of Mathematics and Physics, Center for Particle Physics, Prague, Czech Republic}
\affiliation{Czech Technical University in Prague, Prague, Czech Republic}
\affiliation{Center for Particle Physics, Institute of Physics, Academy of Sciences of the Czech Republic, Prague, Czech Republic}
\affiliation{Universidad San Francisco de Quito, Quito, Ecuador}
\affiliation{LPC, Universit\'e Blaise Pascal, CNRS/IN2P3, Clermont, France}
\affiliation{LPSC, Universit\'e Joseph Fourier Grenoble 1, CNRS/IN2P3, Institut National Polytechnique de Grenoble, Grenoble, France}
\affiliation{CPPM, Aix-Marseille Universit\'e, CNRS/IN2P3, Marseille, France}
\affiliation{LAL, Universit\'e Paris-Sud, CNRS/IN2P3, Orsay, France}
\affiliation{LPNHE, Universit\'es Paris VI and VII, CNRS/IN2P3, Paris, France}
\affiliation{CEA, Irfu, SPP, Saclay, France}
\affiliation{IPHC, Universit\'e de Strasbourg, CNRS/IN2P3, Strasbourg, France}
\affiliation{IPNL, Universit\'e Lyon 1, CNRS/IN2P3, Villeurbanne, France and Universit\'e de Lyon, Lyon, France}
\affiliation{III. Physikalisches Institut A, RWTH Aachen University, Aachen, Germany}
\affiliation{Physikalisches Institut, Universit{\"a}t Freiburg, Freiburg, Germany}
\affiliation{II. Physikalisches Institut, Georg-August-Universit{\"a}t G\"ottingen, G\"ottingen, Germany}
\affiliation{Institut f{\"u}r Physik, Universit{\"a}t Mainz, Mainz, Germany}
\affiliation{Ludwig-Maximilians-Universit{\"a}t M{\"u}nchen, M{\"u}nchen, Germany}
\affiliation{Fachbereich Physik, Bergische  Universit{\"a}t Wuppertal, Wuppertal, Germany}
\affiliation{Panjab University, Chandigarh, India}
\affiliation{Delhi University, Delhi, India}
\affiliation{Tata Institute of Fundamental Research, Mumbai, India}
\affiliation{University College Dublin, Dublin, Ireland}
\affiliation{Korea Detector Laboratory, Korea University, Seoul, Korea}
\affiliation{CINVESTAV, Mexico City, Mexico}
\affiliation{FOM-Institute NIKHEF and University of Amsterdam/NIKHEF, Amsterdam, The Netherlands}
\affiliation{Radboud University Nijmegen/NIKHEF, Nijmegen, The Netherlands}
\affiliation{Joint Institute for Nuclear Research, Dubna, Russia}
\affiliation{Institute for Theoretical and Experimental Physics, Moscow, Russia}
\affiliation{Moscow State University, Moscow, Russia}
\affiliation{Institute for High Energy Physics, Protvino, Russia}
\affiliation{Petersburg Nuclear Physics Institute, St. Petersburg, Russia}
\affiliation{Stockholm University, Stockholm and Uppsala University, Uppsala, Sweden }
\affiliation{Lancaster University, Lancaster LA1 4YB, United Kingdom}
\affiliation{Imperial College London, London SW7 2AZ, United Kingdom}
\affiliation{The University of Manchester, Manchester M13 9PL, United Kingdom}
\affiliation{University of Arizona, Tucson, Arizona 85721, USA}
\affiliation{University of California Riverside, Riverside, California 92521, USA}
\affiliation{Florida State University, Tallahassee, Florida 32306, USA}
\affiliation{Fermi National Accelerator Laboratory, Batavia, Illinois 60510, USA}
\affiliation{University of Illinois at Chicago, Chicago, Illinois 60607, USA}
\affiliation{Northern Illinois University, DeKalb, Illinois 60115, USA}
\affiliation{Northwestern University, Evanston, Illinois 60208, USA}
\affiliation{Indiana University, Bloomington, Indiana 47405, USA}
\affiliation{Purdue University Calumet, Hammond, Indiana 46323, USA}
\affiliation{University of Notre Dame, Notre Dame, Indiana 46556, USA}
\affiliation{Iowa State University, Ames, Iowa 50011, USA}
\affiliation{University of Kansas, Lawrence, Kansas 66045, USA}
\affiliation{Kansas State University, Manhattan, Kansas 66506, USA}
\affiliation{Louisiana Tech University, Ruston, Louisiana 71272, USA}
\affiliation{University of Maryland, College Park, Maryland 20742, USA}
\affiliation{Boston University, Boston, Massachusetts 02215, USA}
\affiliation{Northeastern University, Boston, Massachusetts 02115, USA}
\affiliation{University of Michigan, Ann Arbor, Michigan 48109, USA}
\affiliation{Michigan State University, East Lansing, Michigan 48824, USA}
\affiliation{University of Mississippi, University, Mississippi 38677, USA}
\affiliation{University of Nebraska, Lincoln, Nebraska 68588, USA}
\affiliation{Rutgers University, Piscataway, New Jersey 08855, USA}
\affiliation{Princeton University, Princeton, New Jersey 08544, USA}
\affiliation{State University of New York, Buffalo, New York 14260, USA}
\affiliation{Columbia University, New York, New York 10027, USA}
\affiliation{University of Rochester, Rochester, New York 14627, USA}
\affiliation{State University of New York, Stony Brook, New York 11794, USA}
\affiliation{Brookhaven National Laboratory, Upton, New York 11973, USA}
\affiliation{Langston University, Langston, Oklahoma 73050, USA}
\affiliation{University of Oklahoma, Norman, Oklahoma 73019, USA}
\affiliation{Oklahoma State University, Stillwater, Oklahoma 74078, USA}
\affiliation{Brown University, Providence, Rhode Island 02912, USA}
\affiliation{University of Texas, Arlington, Texas 76019, USA}
\affiliation{Southern Methodist University, Dallas, Texas 75275, USA}
\affiliation{Rice University, Houston, Texas 77005, USA}
\affiliation{University of Virginia, Charlottesville, Virginia 22901, USA}
\affiliation{University of Washington, Seattle, Washington 98195, USA}
\author{V.M.~Abazov} \affiliation{Joint Institute for Nuclear Research, Dubna, Russia}
\author{B.~Abbott} \affiliation{University of Oklahoma, Norman, Oklahoma 73019, USA}
\author{M.~Abolins} \affiliation{Michigan State University, East Lansing, Michigan 48824, USA}
\author{B.S.~Acharya} \affiliation{Tata Institute of Fundamental Research, Mumbai, India}
\author{M.~Adams} \affiliation{University of Illinois at Chicago, Chicago, Illinois 60607, USA}
\author{T.~Adams} \affiliation{Florida State University, Tallahassee, Florida 32306, USA}
\author{G.D.~Alexeev} \affiliation{Joint Institute for Nuclear Research, Dubna, Russia}
\author{G.~Alkhazov} \affiliation{Petersburg Nuclear Physics Institute, St. Petersburg, Russia}
\author{A.~Alton$^{a}$} \affiliation{University of Michigan, Ann Arbor, Michigan 48109, USA}
\author{G.~Alverson} \affiliation{Northeastern University, Boston, Massachusetts 02115, USA}
% ...
%
%
\collaboration{The D0 Collaboration\footnote{with visitors from
$^{a}$Augustana College, Sioux Falls, SD, USA,
$^{b}$The University of Liverpool, Liverpool, UK,
$^{c}$SLAC, Menlo Park, CA, USA,
$^{d}$ICREA/IFAE, Barcelona, Spain,
$^{e}$Centro de Investigacion en Computacion - IPN, Mexico City, Mexico,
$^{f}$ECFM, Universidad Autonoma de Sinaloa, Culiac\'an, Mexico,
and 
$^{g}$Universit{\"a}t Bern, Bern, Switzerland.%
}} \noaffiliation
\vskip 0.25cm

\date{August 12, 2010}

\begin{abstract}
We report a search for diphoton events with large missing transverse energy produced in $p\bar{p}$ collisions at $\sqrt{s}=1.96$~TeV.  The data were collected with the D0 detector at the Fermilab Tevatron Collider, and correspond to 6.3~fb$^{-1}$ of integrated luminosity.  The observed missing transverse energy distribution is well described by the standard model prediction, and 95\%~C.L. limits are derived on two realizations of theories beyond the standard model.  In a gauge mediated supersymmetry breaking scenario, the breaking scale $\Lambda$ is excluded for $\Lambda < 124$~TeV.  In a universal extra dimension model including gravitational decays, the compactification radius $R_{c}$ is excluded for $R_{c}^{-1} < 477$~GeV.
\end{abstract}

\maketitle

In the standard model (SM), events with two high transverse momentum photons ($\gamma \gamma$) and large missing transverse energy (\metb) are produced at a small rate in $p\bar{p}$ collisions.  This final state is therefore sensitive to contributions from processes beyond the SM (BSM).  We report a search for $\gamma\gamma$ events with large \met produced in $p\bar{p}$ collisions recorded using the D0 detector at the Fermilab Tevatron Collider.  The sensitivity is assessed for two benchmark BSM models, gauge mediated supersymmetry (SUSY) breaking (GMSB)~\cite{GMSB} and universal extra dimensions (UED)~\cite{UED}.

In GMSB models, the masses of the SUSY partners to SM particles arise from SM gauge interactions and are proportional to the effective SUSY breaking scale $\Lambda$.  As the gravitino ($\tilde{G}$) does not participate in SM gauge interactions, it has a small mass~\cite{SuperHiggs} and is the lightest SUSY particle.  Assuming $R$ parity conservation~\cite{Rparity}, the SUSY process with the largest cross section at the Tevatron would be chargino and neutralino pair production ({$\chi_{1}^{\pm}\chi_{2}^{0}$, $\chi_{1}^{\pm}\chi_{1}^{\mp}$)~\cite{Prospino}, followed by decay chains to the next-to-lightest SUSY particle (NLSP).  We consider the case when the lightest neutralino ($\chi_{1}^{0}$) is the NLSP~\cite{Martin}, and decays promptly with the dominant branching fraction yielding a photon and an essentially massless gravitino ($\chi_{1}^{0}\rightarrow \tilde{G}  \gamma$)~\cite{Gravitino}.  The two gravitinos escape detection, resulting in the final state $\gamma\gamma +\metmath+X$, where $X$ denotes leptons and jets produced in the decay chains~\cite{GMSBpheno}.

In UED models, extra spatial dimensions are predicted that are accessible to all SM fields.  We consider the case of a single UED that is compactified with radius  $R_{c}$, resulting in a tower of states for each SM field, called Kaluza-Klein (KK) excitations, with the masses of these states separated by $R_{c}^{-1}$.  At the Tevatron, the UED process with the largest cross section would be the production of pairs of first level KK quarks~\cite{UEDprod},  followed by decay chains to the lightest KK particle (LKP), the KK photon ($\gamma^{*}$).  If additional larger extra dimensions also exist that are only accessible to gravity, the LKP is able to decay promptly through gravitational interactions to a photon and a graviton ($\gamma^{*}\rightarrow G  \gamma$)~\cite{KKviolation,UEDpheno}.  The two gravitons escape detection, resulting in the final state $\gamma\gamma +\metmath+X$.

Searches for BSM physics in $\gamma\gamma+\metmath+X$ events have been performed at the CERN $e^+e^-$ Collider (LEP)~\cite{LEP}, and at the Tevatron in Run I~\cite{RunI} and Run II~\cite{CDF1,CDF2,DZERO1,DZERO2}.  This analysis uses similar methods to those adopted in Ref.~\cite{DZERO2}, a six times larger dataset, and improved photon identification criteria utilizing a neural network (NN) discriminant recently employed in other analyses~\cite{d0photon}.  The larger dataset has substantially increased the search sensitivity, and has allowed an improved formulation of the data-derived SM background prediction.  The background prediction, including the assessment of systematic uncertainties, was developed using only the $\metmath \leq 50$~GeV region of the $\gamma\gamma$ sample.  Once finalized, the events with $\metmath>50$~GeV were included in evaluating the consistency with the SM prediction and the sensitivity to the signal models.  In addition to substantially improved limits on the GMSB model, this Letter also presents the first limits on the UED model with gravitational decays.

The D0 detector~\cite{d0det} consists of an inner tracker, a liquid-argon/uranium calorimeter, and a muon spectrometer.  The tracking system is comprised of a silicon microstrip tracker (SMT) and a central fiber tracker (CFT), both located within a 2~T superconducting solenoidal magnet.  A central calorimeter (CC) covers pseudorapidities $|\eta| < 1.1$, and two endcap calorimeters (EC) extend the coverage to $|\eta| <4.2$, where $\eta = -\ln[\tan(\theta/2)]$, and $\theta$ is the polar angle with respect to the proton beam direction.  The electromagnetic (EM) section of the calorimeter is segmented in four longitudinal layers (EM$i$, $i=1,4$) with transverse segmentation $\Delta \eta \times \Delta \phi = 0.1 \times 0.1$ ($\phi$ is the azimuthal angle), except in EM3 where it is  $0.05 \times 0.05$.  A central preshower detector (CPS) utilizing several layers of scintillating strips, positioned between the solenoid coil and CC, provides a precise measurement of EM shower position.  The trajectory of photon candidates is reconstructed by combining the four EM-layer and CPS measurements~\cite{DZERO2}.

The data analyzed were collected with single EM triggers and correspond to an integrated luminosity of $6.3\pm 0.4$~$\mathrm{fb}^{-1}$~\cite{d0lumi}.  Events containing identified calorimeter noise patterns which could bias the \met distribution are removed.  Diphoton candidate events are selected by requiring at least two photon candidates with transverse energy $E_{T}>25$~GeV identified in the CC.  Photon candidates are selected from EM clusters reconstructed within a cone of radius 
${\cal R} \equiv \sqrt{ (\Delta \eta)^{2} + (\Delta \phi)^{2} } = 0.2$ by requiring (i) $\geq 95 \%$ of the cluster energy be deposited in the EM layers, (ii) the calorimeter isolation variable $I \equiv \left[ E_{tot}(0.4) - E_{EM}(0.2) \right]/E_{EM}(0.2)$ be less than 0.10, where $E_{tot}$($\mathcal{R}$) ($E_{EM}$($\mathcal{R}$)) is the total (EM) energy in a cone of radius $\mathcal{R}$ , (iii) the shower width in EM3 be consistent with an EM shower, (iv) the scalar sum of the transverse momentum ($p_{T}$) of tracks originating from the $p\bar{p}$ collision vertex (PV) in a $0.05 < \mathcal{R} < 0.4$ annulus about the cluster centroid be less than $2$~GeV, and (v) the cluster not be spatially matched to a reconstructed track or a significant density of SMT and CFT hits~\cite{DZERO2}.  Further rejection of jets misidentified as photons is achieved with a requirement on the NN discriminant, trained using a set of track, CPS, and calorimeter based variables~\cite{d0photon}.

Electrons satisfy the same requirements as photons, with the exception of the track veto (item v).  Jets are reconstructed with the iterative midpoint algorithm~\cite{d0jets} with cone size $\mathcal{R}=0.5$.  The \met is determined using calorimeter energy depositions with $|\eta|<4$.  Corrections are applied to \met to calibrate energy from EM objects and jets, and to account for the $p_{T}$ of muons.  There are on average several $p\bar{p}$ interactions per crossing of the beams.  The correct PV is identified in $\approx98\%$ of signal events for the benchmark models.  The photon trajectories must indicate that the candidates originate at the PV.  This requirement is to ensure an accurate calculation of transverse energy in background events in which the correct PV is less efficiently identified, to suppress non-collision events, and measured to be $\approx86\%$ efficient using a $Z(\rightarrow ee,\mu\mu)+\gamma$ data sample.  To reduce the number of events with significantly mismeasured \met, events are rejected if the difference in azimuthal angle ($\Delta \phi$) between the highest $E_{T}$ jet (if present) and \met is greater than 2.5 radians, or if $\Delta \phi$ between either photon and \met is less than 0.2 radians.  A total of 7934 $\gamma\gamma$ candidate events satisfy these criteria.

SM background events in the $\gamma\gamma$ sample are categorized as arising from instrumental \met sources (SM $\gamma\gamma$, $\gamma+$jet, multi-jet) and genuine \met sources ($W\gamma$, $W+\mathrm{jet}$, $W/Z+\gamma\gamma$).  All backgrounds are measured using data control samples, with the exception of small contributions from $W/Z+\gamma\gamma$ events, which are estimated using Monte Carlo (MC) simulation.

Instrumental \met is a result of energy mismeasurement in an otherwise $E_{T}$ balanced event.  Instrumental \met sources in the  $\gamma\gamma$ sample are separated into contributions from SM $\gamma\gamma$ events, and events with at least one photon candidate originating from a misidentified jet (misID-jet), {\sl i.e.}, $\gamma+$jet and multi-jet events.  The difference in energy resolution for real photons and fakes from misidentified jets results in a difference in the shape of the \met distribution between the two categories.

The \met shape in SM $\gamma\gamma$ events is modeled using a dielectron ($ee$) data sample predominantly composed of $Z \rightarrow ee$ events.  The $ee$ sample satisfies the same kinematic requirements as the $\gamma\gamma$ sample, with the exception that the $ee$ invariant mass is restricted to an interval about the $Z$ boson peak to reduce genuine \met contributions ({\sl e.g.}, $W+\mathrm{jet}$, di-boson, and $t\bar{t}$ events).  The \met distribution in $ee$ events is compared with shapes in $Z \rightarrow ee$ and SM~$\gamma\gamma$ MC events generated with {\sc pythia}~\cite{pythia}.  These MC samples, and all others used in this Letter, were processed with full {\sc geant}~\cite{geant} detector simulation and standard reconstruction algorithms.  Kinematic differences between the $Z \rightarrow ee$ and SM $\gamma\gamma$ processes are verified with MC to have a negligible impact on the \met shape.  The $Z \rightarrow ee$ MC accurately models $ee$ data for \met values below $\metmath \approx 35$~GeV.  Above this value, a more pronounced tail is observed in $ee$ data.  The tail in data reflects both mismeasurements not modeled in MC, and a small residual presence of genuine \met events in the $ee$ sample. The average of the data and MC shapes is used to model the \met in SM $\gamma\gamma$ events for values of $\metmath > 35$~GeV, and the data-only and MC-only extremes are used to define a systematic uncertainty on the this shape.

The \met shape in misID-jet events is modeled with a data sample satisfying the same requirements as the $\gamma\gamma$ sample with the exception that at least one of the photon candidates fails the NN requirement.  Additionally, photon identification requirements (iii) and (iv) are loosened to reduce the statistical uncertainty on the \met shape.  A systematic uncertainty on the \met shape in events with misidentified jets is obtained by varying the photon identification criteria.

The instrumental \met background estimate is normalized such that the number of events with $\metmath<10$~GeV is equal to that in the $\gamma\gamma$ sample.  The relative contribution of SM $\gamma\gamma$ and misID-jet background events is determined by a fit to the $\gamma\gamma$ sample \met distribution for $\metmath<20$~GeV.  The fit accounts for the small contribution of SM background with genuine \met in the fit region, and is verified to be insensitive to signal contributions for benchmark model cross sections relevant to this analysis.  The SM $\gamma\gamma$ contribution to the $\gamma\gamma$ sample over the full \met range is $(41 \pm 17)\%$.  A systematic uncertainty accounts for changes in the shape of the predicted instrumental \met distribution arising from the uncertainty in the determination of the SM $\gamma\gamma$ contribution.

SM background with genuine \met arises from real SM $\gamma\gamma+\metmath+X$ events and from events with an electron misidentified as a photon (misID-ele). The misID-ele contribution is derived using an $e\hspace{0.01in}\gamma$ data sample, composed primarily of instrumental \met sources for $\metmath<20$~GeV and $W(\rightarrow e\nu)\gamma$ and $W(\rightarrow e\nu)+\mathrm{jet}$ events at higher \met values.  The instrumental \met sources are modeled with the previously introduced $ee$ and misID-jet \met shapes, respectively.  The $Z\rightarrow ee$ normalization is determined by fitting the $Z$ boson peak in the $e\hspace{0.01in}\gamma$ invariant mass distribution, and the multi-jet \met shape is normalized to provide the remaining contribution in the $\metmath < 10$~GeV region. The presence of real \met contributions in the $e\hspace{0.01in}\gamma$ sample is seen as an excess of events with high \met values above the predicted contributions from instrumental sources. This excess is well described by $W\gamma$ and $W+\mathrm{jet}$ events.  The expected $W$ boson peak is observed in the transverse mass distribution of $e\hspace{0.01in}\gamma$ sample events with $\metmath>30$~GeV.  The normalization of the $W+\mathrm{jet}$ contribution is determined from a comparison of the data photon NN shape with MC real and fake photon NN  shapes~\cite{d0photon} in this \met region.  The remaining contribution is in good agreement with {\sc pythia} $W\gamma$ production after applying a next-to-leading order (NLO) QCD correction~\cite{wgamnlo} and an additional $+15\%$ scaling factor accounting for QED final state radiation (FSR) in inclusive $W$ production.  The FSR component~\cite{FSR} is determined with data using the $\Delta \mathcal{R}(e,\gamma)$ distribution.  
The predicted misID-ele contribution to the $\gamma\gamma$ sample equals the excess of high \met events in the  $e\hspace{0.01in}\gamma$ sample, scaled by $f_{e \rightarrow \gamma} /(1-f_{e \rightarrow \gamma})$, where $f_{e \rightarrow \gamma}=0.020\pm0.005$ denotes the rate at which an electron fakes a photon satisfying the selection criteria, as measured with $Z\rightarrow ee$ data.

Real SM diphoton events with large genuine \met originate from $W/Z+\gamma\gamma$ processes.  This small background contribution is estimated with MC using {\sc madgraph}~\cite{MadGraph}.  Events with inclusive $W$ and $Z$ boson decay modes are simulated, with $W \rightarrow l\nu~(l=e,\mu,\tau)$ and $Z\rightarrow \nu\bar{\nu}$ providing the largest genuine \met contribution.  A total of $1.6\pm 0.1$ $W+\gamma\gamma$ events and $3.8\pm 0.3$ $Z+\gamma\gamma$ events are estimated to be present in the  $\gamma\gamma$ sample.  Figure~\ref{fig:ggmet} displays the $\gamma\gamma$ sample \met distribution, which is in good agreement with the SM prediction over the full \met range.  Table~\ref{table:events} provides the observed number of $\gamma\gamma$ sample events and the SM prediction in three \met regions.  

\begin{figure}
\includegraphics[scale=0.43]{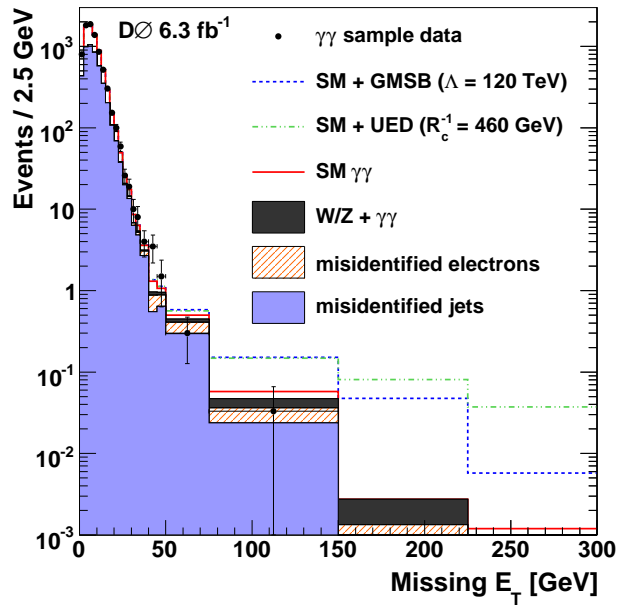}
\caption{\label{fig:ggmet} \met distribution in the $\gamma\gamma$ sample shown with statistical uncertainty and expected SM background from events with a misidentified jet, a misidentified electron, $W/Z+\gamma\gamma$ events, and SM $\gamma\gamma$ events.  The expected \met distribution in the presence of GMSB and UED events is also displayed for example values of $\Lambda$ and $R_{c}^{-1}$, respectively.}
\end{figure}

\begin{table*}[t!]
\caption{\label{table:events} Observed number of $\gamma\gamma$ sample events, predicted background from instrumental \met sources (SM $\gamma\gamma$, $\gamma+$jet, QCD multi-jet) and genuine \met sources ($W\gamma$, $W+\mathrm{jet}$, $W/Z+\gamma\gamma$), and total predicted SM background, in three \met intervals. The expected number of GMSB and UED signal events is listed for two $\Lambda$ and $R_{c}^{-1}$ values, respectively.  The total uncertainty on the SM background and expected signal is given. }
\begin{ruledtabular}
\begin{tabular}{c c ccr  cccc }
\multicolumn{1}{c}{\met Interval,}   & \multicolumn{1}{c}{Observed} & \multicolumn{3}{c}{\text{SM Background Events}}   & \multicolumn{4}{c}{ \text{Expected Signal Events}}  \\ 
\multicolumn{1}{c}{GeV}   & \multicolumn{1}{c}{Events}  & \multicolumn{1}{c}{\text{Instr. \met}} & \multicolumn{1}{c}{\text{Genuine \met}} & \multicolumn{1}{c}{\text{Total}}  & \multicolumn{1}{c}{\text{GMSB}} &  \multicolumn{1}{c}{\text{GMSB}} &  \multicolumn{1}{c}{\text{UED}} &  \multicolumn{1}{c}{\text{UED}}   \\
&   &\multicolumn{1}{c}{\phantom{$\Lambda=100$~TeV}}   &   & \multicolumn{1}{c}{\phantom{$\Lambda=100$~TeV}}  &  \multicolumn{1}{c}{$\Lambda=100$~TeV} &  \multicolumn{1}{c}{$\Lambda=120$~TeV}  &  \multicolumn{1}{c}{$R_{c}^{-1}=420$~GeV}  &  \multicolumn{1}{c}{$R_{c}^{-1}=460$~GeV}   \\    
\cline{1-1}  \cline{2-2} \cline{3-5}  \cline{6-9} 
$35 - 50$  & 18& $9.6\pm1.9$  & $2.3\pm0.5$  & $11.9\pm2.0$ & $\hspace{0.06in}1.8\pm0.1$    & $0.3\pm0.1$    & $\hspace{0.06in}1.4\pm0.1$  & $0.3\pm0.1$  \\
$50 -75$    & 3& $3.5\pm0.8$  & $1.5\pm0.3$  & $5.0\pm0.9$   & $\hspace{0.06in}4.1\pm0.3$    & $0.8\pm0.1$    & $\hspace{0.06in}2.9\pm0.2$  & $0.6\pm0.1$   \\ 
$>75$        &  1& $1.1\pm0.4$  & $0.8\pm0.1$  & $1.9\pm0.4$    & $14.3\pm1.1$                             & $4.4\pm0.4$     & $24.7\pm2.0$   & $6.4\pm0.5$  \\ 
\end{tabular}
\end{ruledtabular}
\end{table*}

We determine the sensitivity to the GMSB scenario using a set of values, termed SPS8~\cite{SPS}, for the model parameters.  In this set the scale $\Lambda$ is unconstrained,  $M_{mes}=2\Lambda$, $N_{mes}=1$, $\tan\beta=15$, and  $\mu>0$~\cite{SPS}.  The masses and decay widths of SUSY particles are calculated with {\sc susyhit 1.3}~\cite{susyhit} and used to generate {\sc pythia} MC events.  The  event selection efficiency is $0.17\pm0.02$ at $\Lambda=120$~TeV, and does not differ significantly for other $\Lambda$ values studied.  The NLO production cross section is calculated with {\sc prospino 2.1}~\cite{Prospino}.  The expected \met distribution for the SM and GMSB at $\Lambda=120$~TeV is depicted in Figure~\ref{fig:ggmet}.  The number of expected GMSB events in three \met regions is listed in Table~\ref{table:events} for $\Lambda=100$ and $120$~TeV.

We consider the UED model as implemented in {\sc pythia 6.421}~\cite{UEDpythia}, leaving $R_{c}^{-1}$ unconstrained and setting $\tilde{\Lambda} R_{c}=20$, where $\tilde{\Lambda}$ is the cutoff scale for radiative corrections to KK masses.  This UED model is implemented in a higher $(4+N)$ dimensional space, where $R_{c}^{-1}$ is much larger than that of the $N$ compact extra dimensions accessible to gravity, inducing KK particle decays through gravitational interactions.  We choose $N=6$ and a fundamental Planck scale $M_{D}=5$~TeV, such that only the $\gamma^{*} \rightarrow G\gamma$ decay occurs with appreciable branching fraction~\cite{UEDpheno}.  The event selection efficiency is $0.19\pm0.02$ at $R_{c}^{-1}=460$~GeV,  and does not differ significantly for other $R_{c}^{-1}$ values studied.  The expected \met distribution for the SM and UED at $R_{c}^{-1}=460$~TeV is depicted in Figure~\ref{fig:ggmet}.  The number of expected UED events in three \met regions is listed in Table~\ref{table:events} for $R_{c}^{-1}=420$ and $460$~GeV.

Systematic uncertainties for sources of instrumental \met are attributed to the uncertainty of the \met shape in SM $\gamma\gamma$ and misID-jet events, and their relative normalization.  An uncertainty in the shape of the \met distribution for the misID-ele contribution arises from the uncertainty in the $Z\rightarrow ee$ contribution to the $e\hspace{0.01in}\gamma$ sample, and a $25\%$ misID-ele normalization uncertainty results from the $f_{e\rightarrow\gamma}$ uncertainty.  Systematic uncertainties in the contributions estimated with MC arise from the integrated luminosity ($6.1\%$), trigger efficiency ($2\%$), and photon identification ($3\%$ per photon) and trajectories ($3\%$) efficiencies.  Uncertainty in parton distribution functions (PDF)~\cite{cteq} yield systematic uncertainties of up to $5\%$ and $20\%$ in the production rate of GMSB and UED events, respectively.

No evidence for BSM physics is observed in the $\gamma\gamma$ sample \met distribution and limits on the benchmark models are derived using a Poisson log-likelihood ratio test~\cite{cls} incorporating the full \met distribution.  Pseudo-experiments are generated according to the background-only and signal plus background hypotheses, and account for statistical uncertainty on the expected number of events and systematic uncertainties.  The cross section limit is evaluated using the CL$_{s}$ modified frequentist approach~\cite{cls}.  Figure~\ref{fig:limits} shows the predicted GMSB and UED cross section with PDF uncertainty, and $95\%$~C.L. cross section exclusion limit, as functions of $\Lambda$ and $R_{c}^{-1}$, respectively.  For GMSB, the NLO cross section uncertainty is small compared to the PDF uncertainty.  The UED NLO cross section has not yet been computed.

In conclusion, we have presented a search for physics beyond the standard model in the $\gamma\gamma+\metmath+X$ final state at the Tevatron.  The observed \met distribution is consistent with the SM expectation and limits on two benchmark models are derived.  In the SPS8 GMSB model, values of the effective SUSY breaking scale $\Lambda < 124$~TeV are excluded at $95\%$~C.L.  The limit excludes $m_{\chi^{0}_{1}}<175$~GeV, representing improvements of 50~GeV~\cite{DZERO2} and 26~GeV~\cite{CDF2} with respect to previous measurements.  Additionally, the first assessment is made of the sensitivity to the UED model with KK particle decays induced by gravitational interactions, excluding values of the compactification radius $R_{c}^{-1} < 477$~GeV at 95\%~C.L. 

\begin{figure*}[]
\includegraphics[scale=0.92]{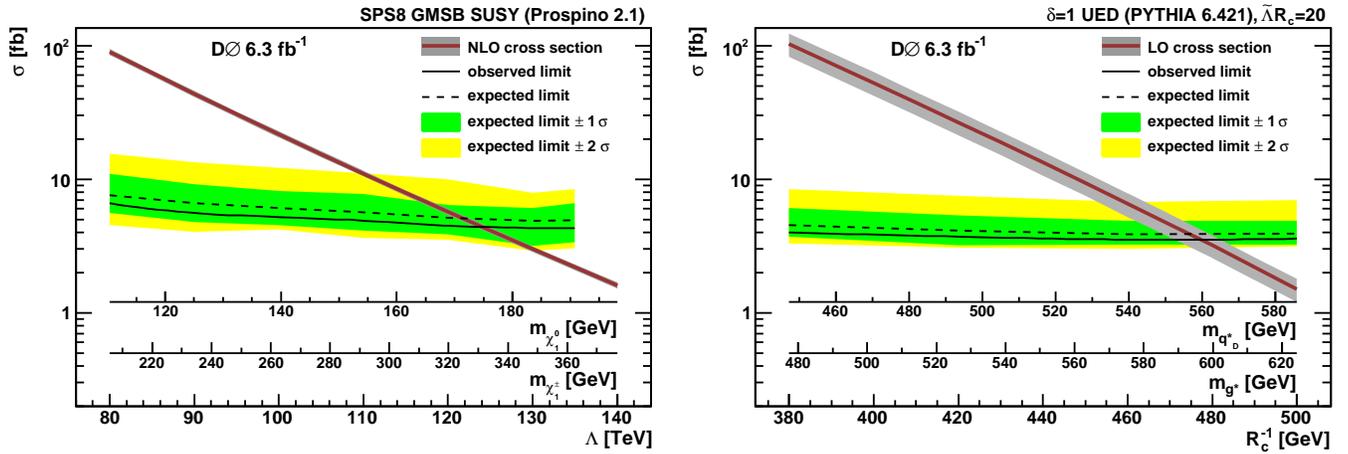}
\caption{\label{fig:limits} The predicted cross section for the benchmark GMSB and UED models, and $95\%$~C.L. expected and observed exclusion limits, as a function of $\Lambda$ and $R_{c}^{-1}$, respectively.  For the GMSB model, corresponding masses are shown for the lightest chargino, $\chi_{1}^{\pm}$, and neutralino, $\chi_{1}^{0}$.  For the UED model, corresponding masses are shown for the KK quark, $q_{D}^{*}$, and KK gluon, $g^{*}$.  The $\gamma^{*}$ mass is approximately equal to $R_{c}^{-1}$.}
\end{figure*}

%
We thank the staffs at Fermilab and collaborating institutions,
and acknowledge support from the
DOE and NSF (USA);
CEA and CNRS/IN2P3 (France);
FASI, Rosatom and RFBR (Russia);
CNPq, FAPERJ, FAPESP and FUNDUNESP (Brazil);
DAE and DST (India);
Colciencias (Colombia);
CONACyT (Mexico);
KRF and KOSEF (Korea);
CONICET and UBACyT (Argentina);
FOM (The Netherlands);
STFC and the Royal Society (United Kingdom);
MSMT and GACR (Czech Republic);
CRC Program and NSERC (Canada);
BMBF and DFG (Germany);
SFI (Ireland);
The Swedish Research Council (Sweden);
and
CAS and CNSF (China).
%

\end{document}